# DISTRIBUTED CLOUD COMPUTING ENVIRONMENT ENHANCED WITH CAPABILITIES FOR WIDE-AREA MIGRATION AND REPLICATION OF VIRTUAL MACHINES

Young-Chul Shim

Department of Computer Engineering, Hongik University, Seoul, Korea
youngcshim@cs.hongik.ac.kr

## ABSTRACT

*When a network application is implemented as a virtual machine on a cloud and is used by a large number of users, the location of the virtual machine should be selected carefully so that the response time experienced by users is minimized. As the user population moves and/or increases, the virtual machine may need to be migrated to a new location or replicated on many locations over a wide-area network. Virtual machine migration and replication have been studied extensively but in most cases are limited within a subnetwork to be able to maintain service continuity. In this paper we introduce a distributed cloud computing environment which facilitates the migration and replication of a virtual machine over a wide area network. The mechanism is provided by an overlay network of smart routers, each of which connects a cooperating data center to the Internet. The proposed approach is analyzed and compared with related works.*

## 1. INTRODUCTION

Cloud computing is a computing paradigm which provides to a large number of users various information technology resources with a high level of scalability using internet technology. In a cloud computing environment users make access to a large scale computing environment through their computing devices connected to the Internet, use necessary information technology resources including applications, storage, operating systems, platforms, etc. as much as they want and at any time that they want, and pay the fee based upon the amount of resources that they have used. From a hardware provisioning and pricing point of view, a cloud computing environment has advantages as follow[1]:

- The appearance of infinite computing resources available on demand, quickly enough to follow load surges, thereby eliminating the need for cloud computing users to plan far ahead for provisioning.
- The elimination of an up-front commitment by cloud users, thereby allowing companies to start small and increase hardware resources only when there is an increase in their needs.
- The ability to pay for use of computing resources on a short-term basis as needed (for example, processors by the hour and storage by the day) and release them as needed, thereby rewarding conservation by letting machines and storage go when they are no longer useful.

The places where cloud computing resources are installed and provided to remote users are called data centers. One cloud company can have one data center or many data centers distributed over a wide-area network. Currently there are many cloud companies and their data centers are operated either independently or in a federated mode.

To take the advantage of cloud computing, more and more application service providers run their application programs at data centers and application service users access them across a network. When a cloud-based application service is initially installed on a data center, the location of the data center should be carefully selected so that it can be best suited to the current population of users in such a way that the response time experienced by users is minimized. While the cloud-based application service is being provided, the user population can change. In the first case the location where a majority of users are can change and these majority users will experience excessive response time. In this case it will be better to find a new data center location which is closer to the current user population and move the application service to this new data center. In the second case the cloud-based application service becomes very popular and its user population increases at many locations over the world. In this case it will be better to select many data centers and replicate the application service at these selected data centers.

An application program is executed as a virtual machine(VM) on a cloud and current cloud computing environments provide mechanism to migrate and replicate VMs. To avoid the service disruption, the VM maintains the same IP address after a migration. Thus in general VM migration is limited within a subnetwork or at most within a single data center and VM migration across data center boundaries in a wide-area network remains a difficult problem. When an application service is replicated on many VMs over a wide-area network, users expect that these VMs will have the same IP address so that they can invoke the application service with the same IP address from any location. But current study of VM replication is limited within a subnetwork or a single data center[2].

In this paper we introduce a distributed cloud computing environment which facilitates the migration and replication of VMs over a wide area network. The distributed cloud computing environment consists of many data centers distributed over a wide-area network. These data centers may belong to one cloud company or they may belong to many different companies and be operated in a federated mode. Each data center is connected to the Internet through a smart router. These smart routers form an overlay network among them in the Internet. Using this overlay network any smart router can notify an event occurrence to all the other smart routers of federated data centers. An application service is implemented as a VM at some data center. Any VM which is expected to be migrated or replicated in the future receives an IP address from an anycast address block allocated to the whole federated data centers, when it is created. When this VM is migrated or replicated, the migration to or replication at the destination data center is notified to all the smart routers. Therefore, the smart routers always know the current location of migrated or replicated VMs. When a user sends a message to a migrated/replicated VM, the message is delivered to the closest smart router. Because the smart router knows all the current locations of the invoked VM, it forwards the message to the data center to which the VM has been migrated or to the closest data center among all the data centers at which the VM is replicated. The proposed method allows users of cloud-based application services to experience satisfactory response time although the user population of service users move locations or increases/decreases.

The rest of the paper is organized as follows. We first describe the overall architecture of the proposed system in Section 2 and then explain how VM migration and replication are supported in the proposed architecture in Section 3. In Section 4 we analyze the proposed architecture and compare with related works. Concluding remarks are in Section 5.

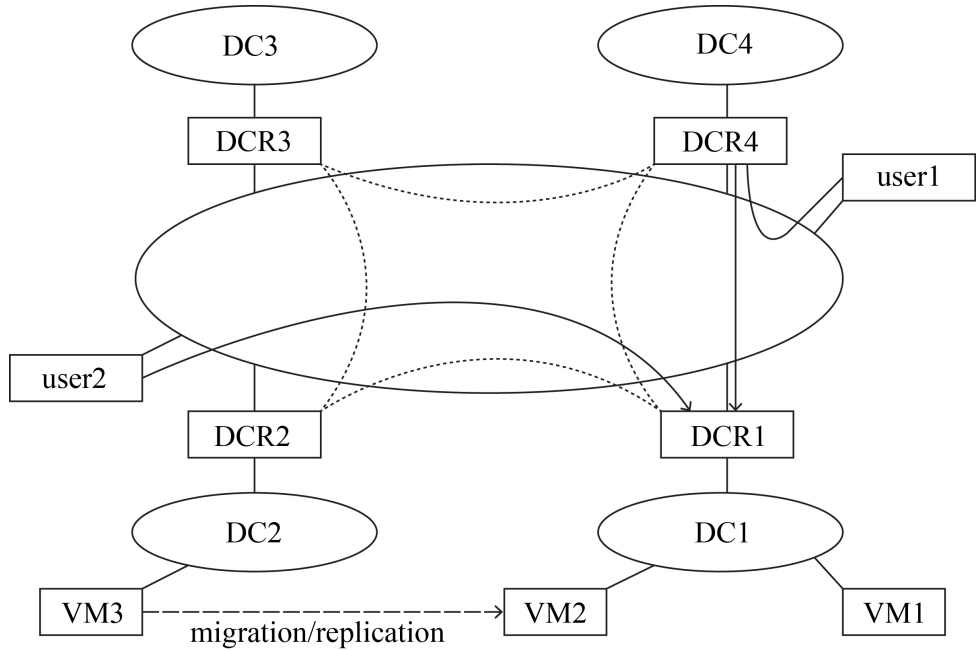

Figure 1. Overall Architecture

## 2. OVERALL ARCHITECTURE

In this section we describe the overall architecture of the proposed distributed cloud computing environment. Figure 1 depicts the overall architecture. In the figure each data center is connected to the Internet through a smart router called DCR (Data Center Router). Four data centers are shown here. All of them may belong to one cloud company or they may belong to different companies but cooperate in a federated mode.

Data centers receive two kinds of address blocks: a unicast address block and an anycast address block. To each data center its own unicast address block is allocated. If an application service provider expects that the application service will not need to be migrated or replicated across the data center boundary, the application service is installed as a VM at a selected VM with a unicast address of that data center. If an application service user sends a packet to the VM, the packet is directly sent to the data center containing that VM by the normal routing mechanism of the Internet.

For all the federated data centers a large block of an anycast addresses is allocated. This big anycast address block is divided into subblocks, and each subblock is assigned to each data center. If an application service provider expects that the VM for the application service is started on a certain data center but will be migrated to or replicated on another data center across the wide area network, the application service is installed as a VM at a selected data center with an anycast address of that data center. We assume that every DCR knows which anycast address subblock is assigned to which data center. But every DCR advertises to the Internet using a routing protocol such as BGP as if the whole anycast address block is assigned to its own data center. Therefore if an application service user sends a message to a VM with an anycast address, the message is delivered to the closest DCR and it forwards the message to the actual data center which contains the invoked VM.

DCRs form an overlay network among them as dotted lines in Figure 1. If a VM migrates from one data center to another data center, the DCR of the destination data center notifies the VM migration to all the other DCRs through the overlay network. If a VM which was originally installed at one data center is replicated at another data center, the DCR of the new data center notifies the VM replication to all the other DCRs through the overlay network. The overlay network is built as a partially connected mesh among DCRs. If a DCR has any message to be broadcast to all the DCRs, the message is flooded over the overlay network. Therefore, when a VM is migrated to or replicated at a new data center, its new location is known to all the DCRs. Using this information, all the DCRs can correctly forward any packets going to a VM with an anycast address.

In Figure 1, there are 4 data centers, each of which is connected to the Internet through a DCR. The data center DC1 has VM1 which was started with a unicast address and, therefore, cannot be migrated or replicated outside DC1. The data center DC2 has VM2 which had been started with an anycast address of DC2 and VM2 was migrated to or replicated at DC1. If user1 wants to send a packet to VM1, the packet is delivered directly to DC1. If user2 wants to send a packet to VM2, the packet is first delivered to DCR4 which is the closest DCR from user2 and then is forwarded to DCR1 and then DC1 because DCR4 knows the current location (in case of migration) or the closest location (in case of replication) of VM2.

## 3. VM MIGRATION AND REPLICATION

In this section we describe how a VM is migrated or replicated in a wide-area network and how packets are exchanged between a user and a migrated or replicated VM. Subsection 3.1 discusses VM migration and subsection 3.2 explains VM replication.

### 3.1. VM Migration

We explain VM migration with an example shown in Figure 2. Initially a VM is created at the data center DC1 and then migrated to the data center DC2. We assume that DCR4 is the closest DCR from the user.

When the VM is created at DC1, it is created with an anycast address allocated to DC1. When the user sends a message to the VM, the message has this anycast address as a destination IP address. But because every DCR advertises to the outside routers as if the whole anycast address block for all the federated clouds is assigned to the data center that is connected to itself, the message sent from the user to the VM at DC1 is first sent to DCR4. By this time, the VM has not migrated yet and DCR4 has not received any report on VM migration. DCR4 checks the destination address of the message and finds that the address belongs to DC1. We assume that every DCR knows which anycast address belongs to which data center. This is possible because when data centers are configured to operate in a federated mode, each DCR can notify to other cooperating DCRs which anycast address subblock is assigned to its data center. DCR3 encapsulates the message from the user with another IP header and forwards it to DCR1. The source address of the outsider IP header is DCR4's address and the destination address is DCR1's address. Therefore, the message is forwarded to the VM at DC1. A reply message from the VM to the user can be directly delivered without IP tunnelling.

Now the VM migrates from DC1 to DC2. After the VM is migrated to DC2, DCR2 notifies the VM migration to all the DCRs through the overlay network among DCRs. The notification message contains the event type (VM migration), the anycast address of the VM, and the IP address of DCR2. We assume that all the notification messages are also sent back to the sender itself. Upon receiving this message, all the DCRs store this information in their VM forwarding table as in Figure 2. Now the user sends a message to the VM and the message is sent to DCR4 as before migration. Upon receiving the message, DCR4 checks its VM forwarding table and finds that the VM has migrated to the data center which is connected to DCR2. DCR4 forwards the message using IP tunnelling but this time the destination address of the outside IP header is the address of DCR2 not DCR1. Therefore, the message is delivered to DC2. A reply message from the VM to the user can be directly delivered without IP tunnelling as before.

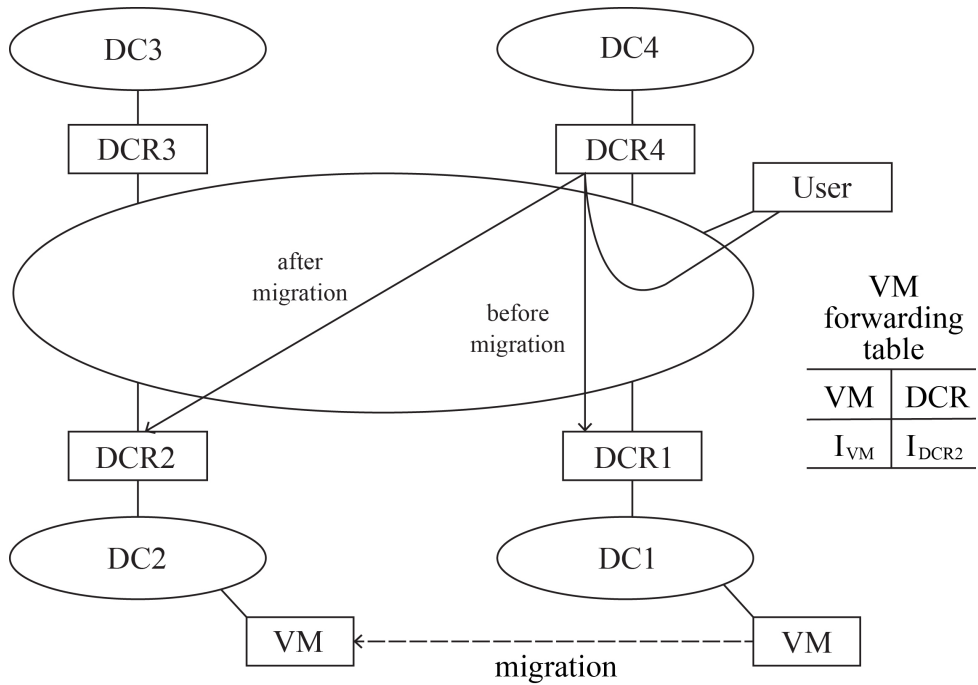

Figure 2. Virtual Machine Migration

When the VM is destroyed at DC2, the VM destruction notification message is broadcast from DCR2 to all the DCRs. The VM destruction message contains the event type (VM destruction), the VM's anycast address, and the address of the DCR of the DC at which the VM is destroyed. All the DCRs receiving this notification message delete the corresponding VM forwarding table entry for this VM.

### 3.2. VM Replication

We explain VM replication with an example shown in Figure 3. In this subsection we consider a scenario that initially a VM is created at the data center DC2 and then replicated at the data center DC3. We assume that DCR4 is the closest DCR from the user and the DCR3 is closer from DCR4 than DCR2 is.

When the VM is created at DC2, it is created with an anycast address allocated to DC2. So the message that a user sends to the VM is delivered to DCR4 and then forwarded to DCR2 as explained the previous subsection. By this time a VM forwarding table entry for this VM is not created in any DCRs yet.

Now we assume that the VM is replicated at DC3. Of course, this replicated VM uses the same anycast address initially allocated at DC2. After the VM is replicated at DC3, DCR3 notifies the VM replication to all the DCRs through the overlay network among DCRs. The notification message contains the event type (VM replication), the anycast address of the VM, the IP address of DCR2, and the IP address of DCR3. Here DCR2 is the replication source and DCR3 is the replication destination. Upon receiving this message, all the DCRs store this information in their VM forwarding table as in Figure 3. Note that in the table the DCR addresses for the replicated VM has both the replication source and replication destination. If we include only the replication destination in the VM forwarding table, all DCRs will misunderstand that the VM is at DC3 but not at DC2. In some cases the replication source can be already stored in the VM forwarding table and, of course, in this case the replication source does not need to be stored in the table again. Now the user sends a message to the VM and the message is sent to DCR4 as before replication. Upon receiving the message, DCR4 checks its VM forwarding table and finds that the VM is located at DC2 and DC3. Between these two data centers, DCR4 chooses DC3 because it knows that DC3 is closer from DCR4 than DC2 is. Therefore, the message is forwarded to the VM at DC3 via DCR3 using IP tunnelling. A reply message from this replicated VM is directly delivered to the user without using IP tunnelling.

When a VM is destroyed at either DC2 or DC3, the VM destruction notification message is broadcast to all the DCRs. The VM destruction message has the same format as in the destruction of a migrated VM. Upon receiving this notification message, the DCR address is deleted from the corresponding VM forwarding table entry. If the DCR address is empty, the VM forwarding table entry for that VM is deleted from the table.

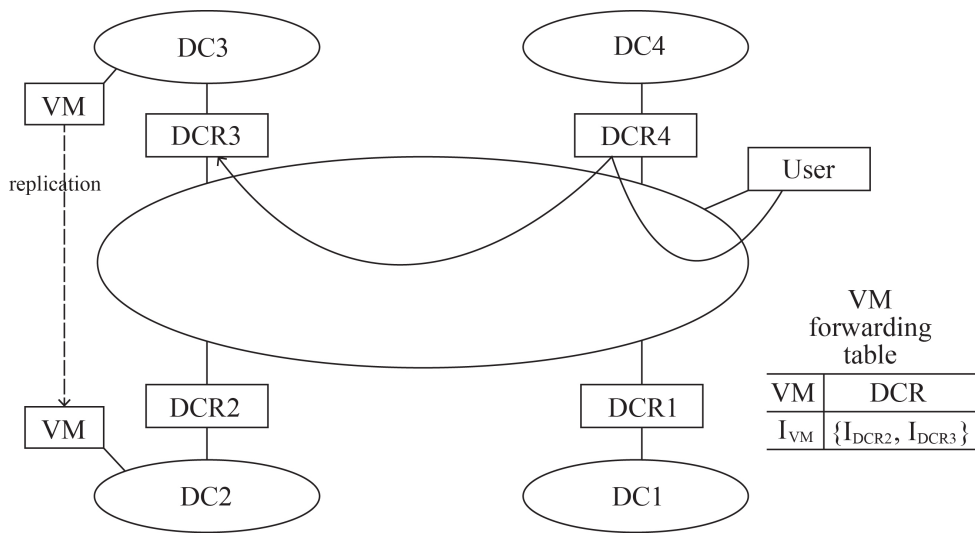

Figure 3. Virtual Machine Replication

## 4. ANALYSIS OF THE PROPOSED ARCHITECTURE

In this section we discuss the performance of the proposed architecture. We then propose algorithms for building an overlay network among data center routers and compare the performance of proposed algorithms through simulation.

### 4.1. Performance of the Proposed Architecture

In this subsection we analyze the message transmission delays in the proposed architecture and also discuss its limitations.

When a VM is not to be migrated or replicated across data center boundaries, it receives a unicast address from the data center at which it is deployed. Therefore, messages exchanged between a user and the VM are directly delivered through the native message forwarding mechanism of the Internet and do not suffer from any degradation in the message transmission delay.

When a VM is expected to be migrated or replicated across data center boundaries, it receives an anycast address from the data center at which it is initially deployed. It retains the same address while it is migrated or replicated. When a user sends a message to this VM, the message is first delivered to the closest DCR and then forwarded to the proper data center. This indirect delivery of messages from the user increases the message delivery delay. But if we have enough number of data centers and deploy them at proper locations, the distance between the user and the closest DCR will be close enough and the increase in the message delivery delay due to the indirect delivery will not be a serious problem. Moreover, the reply messages from the VM to the user is directly delivered to the user and do not experience additional delay. The packets from the user to the VM are delivered using IP tunnelling and, therefore, the size of packets becomes longer due to the added outside IP header and these longer packets can be fragmented during delivery. Performance degradation due to this IP tunnelling cannot be avoided, but we believe that benefits obtained from this IP tunnelling fully outweigh the performance losses due to IP tunnelling.

Now we discuss the issue of service continuity after migration or replication. VM migration in a wide-area network does not have service disruption problem. We explain this with Figure 2. Let us assume that the user started a TCP session with the VM while the VM was located at DC1. After the VM is migrated to DC2, the TCP session between the user and the VM is maintained. It is because when the VM is migrated to DC2, the TCP session information with the user is also check-pointed and then transferred.

But VM replication in a wide-area network can have service discontinuity problem. We explain this with Figure 3. Let us assume that the user started a TCP session with the VM at DC2 before it is replicated. After the VM is replicated at DC3, DCR4 forwards the message from the user to the VM at DC3 because DCR3 is closer from DCR4 than DCR2 is. When the VM is replicated at DC3, it is started without any information about the on-going TCP sessions of the VM at DC2. Therefore, any TCP packets to the VM at DC3 from the user will be rejected and replied with a TCP reset signal. So the user suffers from the service discontinuity problem. We can fix this problem by making the smart router DCR4 remember the unfinished TCP session between the user and the VM at DC2. With this information if a packet from the user to the VM at DC2 arrives at DCR4, DCR4 can forward this packet to the VM at DC2 instead of the VM at DC3. To achieve service continuity using this mechanism requires the smart routers to remember states of all the on-going TCP sessions. This requirement makes the smart router work as a stateful layer 4 switch and imposes excessive performance burden on the smart routers. Due to this problem we do not insist on providing service continuity in VM migration over a wide-area network. When this service discontinuity problem occurs, the user just can start the session with a VM at a new location. But we think that the service discontinuity problem will not occur frequently.

If we want to provide VM migration and replication together, the service discontinuity problem explained above will be aggravated. Thus we do not consider providing VM migration and replication together in this paper.

### 4.2. The Overlay Network among Data Center Routers

In this subsection we discuss the issue of building an overlay network among DCRs. This overlay network is used among DCRs to distribute the location of a migrated or replicated VM. Therefore, this overlay network should be built in such a way that the worst case packet delay among DCRs is minimized and the overhead due to flooding a notification packet in the overlay network should also be minimized.

To compare various strategies for an overlay network, we developed three algorithms for building an overlay network and compared their performance with simulation. To explain and compare the algorithms, we assume that data center routers are located as in Figure 4 (a). To make the figure simple we do not include other routers and interconnections among routers in the figure. Before running the overlay network building algorithm, we assume that all DCRs know the location of all the other DCRs and minimum distance path to them.

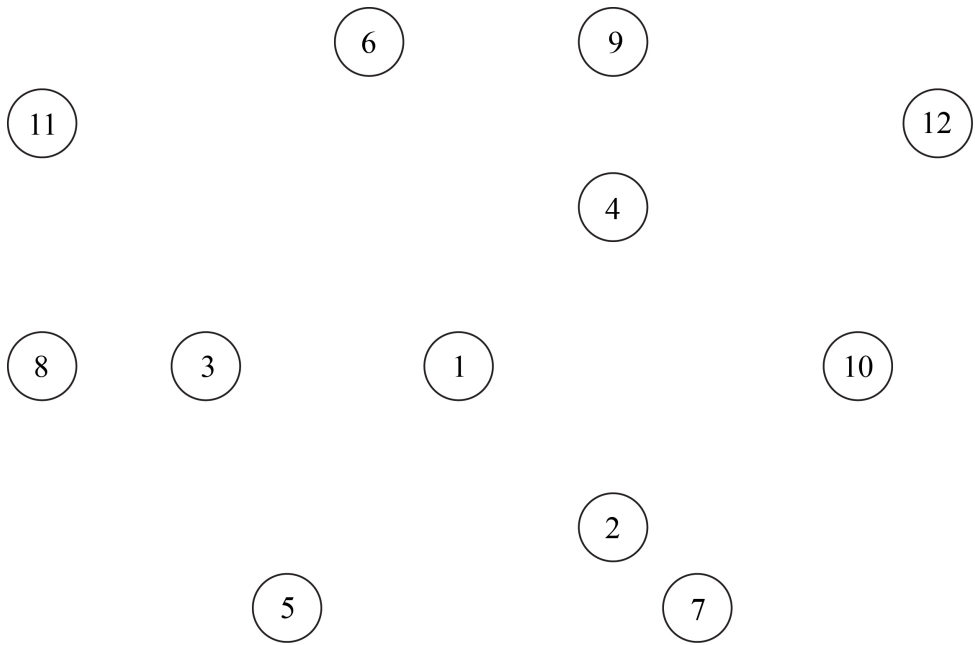

(a) Locations of Data Center Routers

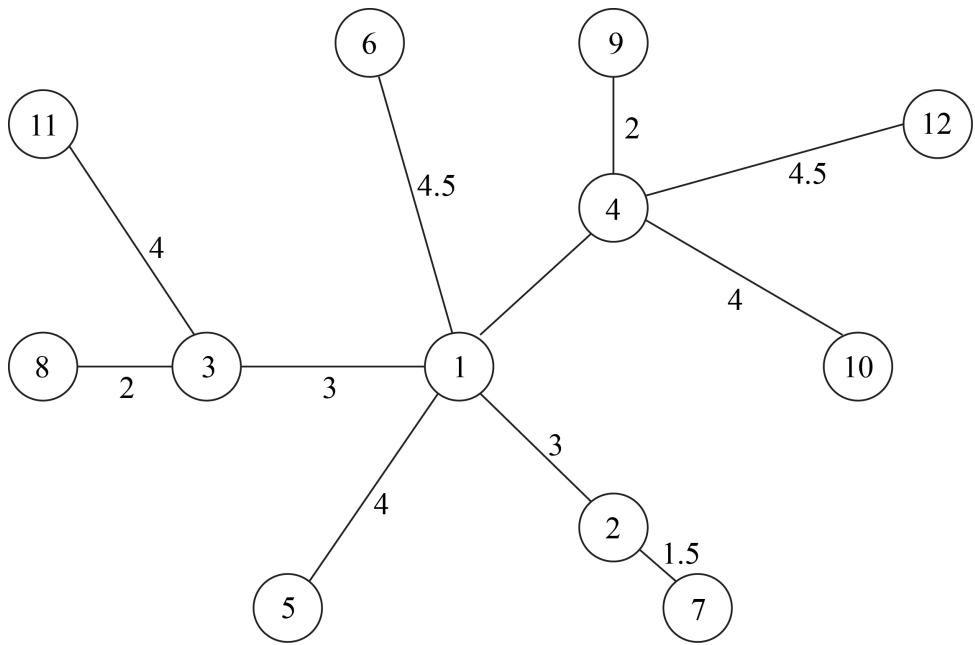

(b) The Result of Algorithm 1

Figure 4. Simulation Results for Overlay Network Construction

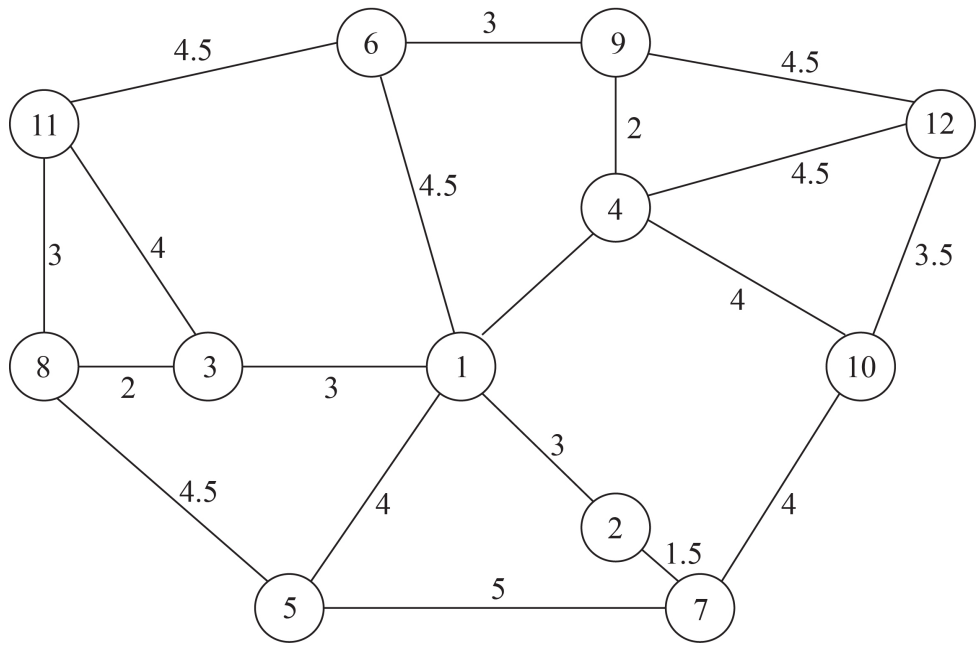

(c) The Result of Algorithm 3

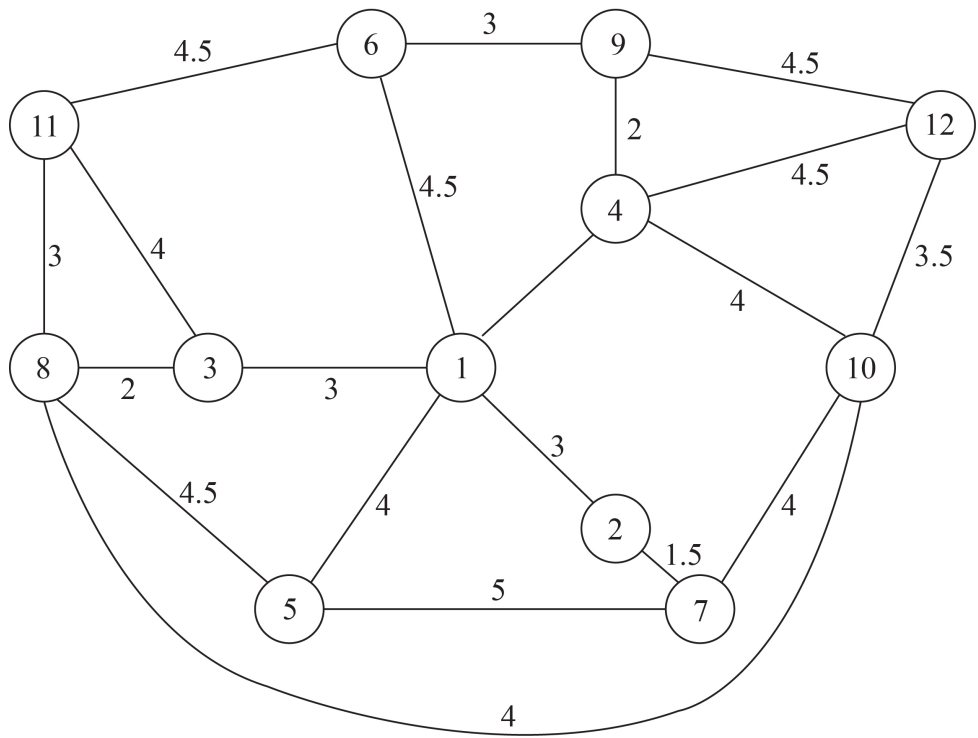

(d) The Result of Algorithm 4

Figure 4. Simulation Results for Overlay Network Construction - Continued

The first algorithm builds a tree among DCRs. The DCR which is closest from the center of the map is selected as the root and other DCRs are added to the tree one by one by a greedy method to build a spanning tree covering all the DCRs. The tree building algorithm is as follow:

```
DCR(1) is selected as the root;
N = the total number of DCRs;
DCR_SET = the set of DCRs except the root;
TREE_SET = {DCR(1)};
sort DCRs in DCR_SET in the increasing order of the distance from the root;
for (i=1; i<=N-1;i++) {
        DCR(j) = the first DCR in DCR_SET;
        from TREE_SET select the closest DCR, DCR(k), from DCR_SET;
        if (DCR(k)) = root)
                connect DCR(j) to DCR(k)
        else {
                find the parent DCR, DCR(m), of DCR(k) on its path to the root;
                DIRECT_COST = the link cost of the direct path
                        from DCR(j) to DCR(m);
                INDIRECT_COST = the total link cost of the indirect path
                        from DCR(j) to DCR(m) via DCR(k);
                if (INDIRECT_COST is at least 25% higher than DIRECT_COST)
                        connect DCR(j) to DCR(m)
                else
                        connect DCR(j) to DCR(k);}
        remove DCR(j) from DCR_SET and include it in TREE_SET;}
```

The algorithm adds DCRs to the tree in the increasing order of their distance from the root. When a certain DCR is added to the tree, it is basically connected to the closest DCR already included in the tree. But if the connection to the parent DCR of the closest DCR gives a shorter path, it is connected to the parent DCR. Figure 4(b) shows the resulting tree. The numbers on edges represent link cost, which is defined to be link delay.

If the overlay network is built just as a tree, the packet delay between neighbouring leaf node DCRs can become very high. For example, the packet from the node 10 to the node 6 should go through the root in Figure 4 (b) although there can be a much shorter path between them. To solve this problem, the second algorithm connects all the leaf node DCRs as in Figure 4 (c).

Because we draw the network map on a plane, the left side looks far away from the right side. But because the earth is globe, the left side should meet the right side. To take this fact into consideration, in the third algorithm, we select the node closest from the left side center and the node closest from the right side center and connect them. The result is shown in Figure 4 (d).

We compared the quality of overlay networks produced by three algorithms with simulation. The performance indices are the worst delay, and the average delay, and the flooding overhead. The flooding overhead is defined as the sum of all link cost paid to flood a notification message on the overlay network. The result is summarized in Table 1. The algorithm 1 incurs the least overhead but its delay figures are the highest. Algorithms 2 and 3 have lower delay figures but pay 160%~170% higher cost. The algorithm 3 improves the delay figures of the algorithm 2 a lot but the cost increase is negligible. Although the algorithm 1 generates a tree which requires least overhead, but because the failure of a single node or link on the tree partitions the tree, it is not proper to use a tree as an overlay network. We choose to use the overlay network generated by the algorithm 3.

Table 1. Simulation Results

|  | Worst Delay | Average Delay | Flooding Overhead |
|---|---|---|---|
| Algorithm 1 | 14.5 | 9.86 | 35.5 |
| Algorithm 2 | 12.5 | 8.23 | 93 |
| Algorithm 3 | 11.5 | 7.55 | 97 |

### 4.3. Comparison with Related Works

Migration of a VM has been extensively studied by many researchers[3]. But to be able to maintain the service continuity, in most cases the migration is limited within a subnetwork so that the migrated VM can use the same IP address which was assigned before migration. Bradford et al studied the issue of live wide-area migration of VMs[4]. Noting that VM migration within a subnetwork focuses on transferring the run-time memory state of the VMs in a subnetwork, they introduce a mechanism to transfer VM's local persistent state (its file system). The VM gets a new IP address after migration. To solve the service discontinuity problem, they propose a temporary network redirection scheme to allow the VM's open network connections to continue even after migration and an announcement of new addresses via dynamic DNS to ensure that new connections are made to the new IP address. But because packets coming from the open connections are first sent to the old location of the VM and then redirected to the new location, they suffer from the performance degradation due to triangular routing. New connections can get a new address from DNS servers which are dynamically updated to announce the new address. But it is well known that due to the caching effect of DNS, many users wishing to get the new address may get the old address and, therefore, fail to get access to the migrated VM.

IP mobility can be used to support VM migration. With mobile IP, when a VM is migrated, it continues to use the same IP address. The router of the migration source network stores the information on the migration destination network and forwards any packets to the VM to the new location using IP tunnelling. But this scheme increases the packet delay due to triangular routing[5]. To solve this problem, route optimization scheme has been proposed[6]. When a VM is migrated, the information on the destination network is informed to the users of that VM. With this information, they can send packets directly to the new location using IP tunnelling. But we see think it unlikely that IP mobility including the route optimization scheme will be deployed widely in the near future. Nevertheless we compare the delays of packets from clients to clouds for two cases: IPv6 with route optimization and our proposed method. Even with IPv6 the VM replication problem cannot be solved. So we consider only the VM migration. From the figure 4 (d) we assume that a client is located between DCRs 5 and 7 and the VM is migrated to the DCR 6. The packet delay is 8.5 in case of IPv6 and 10 in case of our system. If we assume that a client is located between DCRs 7 and 10 and the VM is migrated to the DCR 11, then the packet delay is 12.5 in IPv6 and 13.5 in our system. So we can see that the increase of packet delays in our system due to the indirect packet delay through nearest DCR is not excessive (less than 20% in our simulation cases) compared with the IPv6 with route optimization.

Hao et al proposed a new approach which enables VM migration across multiple networks without losing service continuity[7]. Each data center is connected to the Internet through a smart router called FE(Forwarding Element). All the FEs form a virtual network over the Internet and can be built on SoftRouter[8] and Openflow[9]. When a VM is migrated from a one data center to another data center, the migration event is notified to the CC(Centralized Controller), which then notifies the new location of the VM to all the FEs. When a user sends a packet to the VM, the packet is sent to the nearest FE, which forwards the packet to the migration destination data center. Their approach seems quite similar to our approach as far as VM migration is concerned but has many drawbacks. Because all the packets are delivered through the virtual network among FEs, the packets going to the VM which does not need to be migrated should be delivered indirectly and thus suffers from the packet delay increase. All the migration events are notified to CC, therefore, CC becomes the performance and fault-tolerance bottleneck. Moreover, their approach does not support VM replication at all.

Mechanisms to replicate VMs and use them to speed up computation have been studied by many researchers[10]. But the replication range is limited within a subnetwork and efforts to replicate VMs on data centers located over a wide-area network and provide users a way to access to the nearest VM can be rarely found.

## 5. CONCLUSIONS

When a network application is implemented as a virtual machine on a cloud and is used by a large number of users, the location of the virtual machine should be selected carefully so that the response time experienced by users is minimized. As the user population move or increases, the virtual machine may need to be migrated to a new location or replicated at many locations in a wide-area network. The subject of virtual machine migration and replication has been studied extensively but in most cases is limited within a subnetwork to be able to maintain service continuity. In this paper we introduced a distributed cloud computing environment which facilitated the migration and replication of a virtual machine over a wide area network. The mechanism is provided by an overlay network of smart routers (called data center routers), each of which connects a cooperating data center to the Internet. When a virtual machine which implements a network application is created and is expected to be migrated or replicated across data center boundaries, it is started with an anycast address. Whenever it is migrated or replicated, this event is notified to all the data center routers using the overlay network and, therefore, all the data center routers know the current location or all the replication locations of the virtual machine. When a user sends a packet to the virtual machine, data center routers send the packet to the virtual machine's current location or closest location. The proposed approach was analyzed and it was shown that it provided the capability of virtual machine migration and replication in a wide-area network without causing too much increase in packet delays. Algorithms for building the overlay network among data center routers were introduced and their performance was compared.

## ACKNOWLEDGEMENTS

This work was supported by 2012 Hongik University Research Fund.## REFERENCES

[1]   Armbrust, M. et al, (2010) "A View of Cloud Computing", *Communications of the ACM*, Vol. 53, No. 4, pp.50-58.

[2]   Azodolmolky, S. et al, (2013), "Cloud Computing Networking: Challenges and Opportunities for Innovations," *IEEE Communications Magazine*, Vol. 51, No. 7, pp.54-62.

[3]   Clark, C. et al (2005) "Live Migration of Virtual Machines", *2nd Symposium on Networked Systems Design & Implementation*, pp. 273-286.

[4]   Bradford, R., Kotsovinos, E., Feldmann, A., & Schioeberg, H., (2007) "Liver Wide-Area Migration of Virtual Machines Including Local Persistent State", *Proceedings of the 3rd International Conference on Virtual Execution Environments*, pp. 169-179.

[5]   Perkins, E.C., (2002) "RFC3344: IP Mobility Support for IPv4", www.ietf.org.

**Authors**

Young-Chul Shim received a Ph.D in computer science from University of California, Berkeley. He is currently a professor of the Department of Computer Engineering at Hongik University. His research areas include wireless and mobile networks, Internet protocols, and cloud computing.